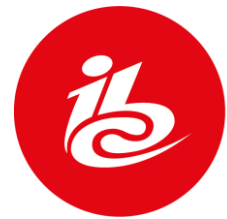

# IMMERSIVE EXPERIENCES AND XR: A GAME ENGINE OR MULTIMEDIA STREAMING PROBLEM?

Simon N.B. Gunkel, Emmanouil Potetsianakis, Tessa E. Klunder,

Alexander Toet, Sylvie S. Dijkstra-Soudarissanane

TNO, Netherlands

**ABSTRACT**

Recent improvements in Extended Reality (XR) technology created an increase in XR products and solutions in the industry, while raising new requirements for new or improved architectural concepts. This need can be particularly complex as XR applications often relate both to 3D geometric rendering and multimedia paradigms. This paper outlines the main concepts relevant to XR, both from a game engineering and multimedia streaming system perspective. XR requires new metadata and media/game orchestration to allow complex interaction between users, objects, and (volumetric) multimedia content, which also results in new requirements on synchronisation (i.e., for global object state and positioning). Furthermore, the paper presents the functional blocks needed in new XR system architectures and how they will glue both (game and media) spaces together. The discussion of functional components and architecture relates to the ongoing activities in relevant standardisation bodies like Khronos, MPEG, and 3GPP. To make XR successful on the long term, the industry needs to agree on interoperable solutions and how to merge both game and media paradigms to allow complex multi-user XR applications.

## INTRODUCTION

While recent advances in AR and VR technology create an uprise in XR and immersive applications, still many open technical issues remain to be solved by the industry. One core aspect of immersive applications or XR technology is to show content in virtual spaces. The user experiences a virtual space and can interact with different objects or media elements. Decades of research and innovation work in the game industry focused on exactly this topic: from the early days and first virtual experiences (e.g., DOOM) towards recent video games with complex lighting like ray tracing (e.g., Cyberpunk 2077). A different direction in XR is to bring photorealistic multimedia streams into the immersive space (i.e., 360-degree and volumetric media). However, when merging these two paradigms in XR, technological gaps and paradigm conflicts can hinder a widespread development of products and solutions.

XR is a superset that covers all aspects related to AR, VR, MR, and its combinations, including technologies that create completely virtual experiences up to experiences that completely blend into reality. It includes cross-interactions between humans, machines, and robots with a promise of a higher immersion and (social) presence than other experience technologies. While there are also many recent advances on human-computer-interaction and quality of experience relating to XR, this paper's focus is on the technical requirements and XR architecture.

VR has been extensively studied since the 1950s (1) with the invention of many different input and output devices. Within the last decade, a new wave of technology advances in the field of AR and VR led to a significant uptake of new solutions in the market, where XR is placed as an important technology for multiple industry verticals (e.g. entertainment, education, healthcare and training). An overview of the most relevant multi-user and collaborative XR use cases can be found in (2). The current progress and success of XR can also be accredited to many ongoing activities in recent years in key industry and standardisation bodies such as MPEG, 3GPP, VRIF, and Khronos.

In particular, we highlight two technical requirements that come with the new wave of XR applications and frameworks, which are a) Spatial Computing and b) Remote (or Split) Rendering. a) Having an understanding of the world, the user's behaviour, as well as spatial properties of objects and multimedia assets is essential for the rendering of virtual (or hybrid) environments and the interaction therein (Spatial Computing). This also comes with the availability of more powerful computing resources and many advances in image analysis that allow an increasing number of Spatial Computing use cases. A simple example in multimedia is background replacement or blurring of the users' webcam feed, currently provided by most video conferencing solutions. Furthermore, we can also see many changes in modern video production chains that now often include game engines, like in the recording of Mandalorian (3). b) Making XR accessible on lightweight AR glasses with low battery consumption, raises the need to move computation from the end device to the cloud (or edge). Fortunately, cloud gaming is already a very established concept that allows remote access of high-powered gaming computations on low powered devices. This can be a good start for XR remote rendering, but with more stringent requirements some changes in multimedia and gaming architectures need to be considered.

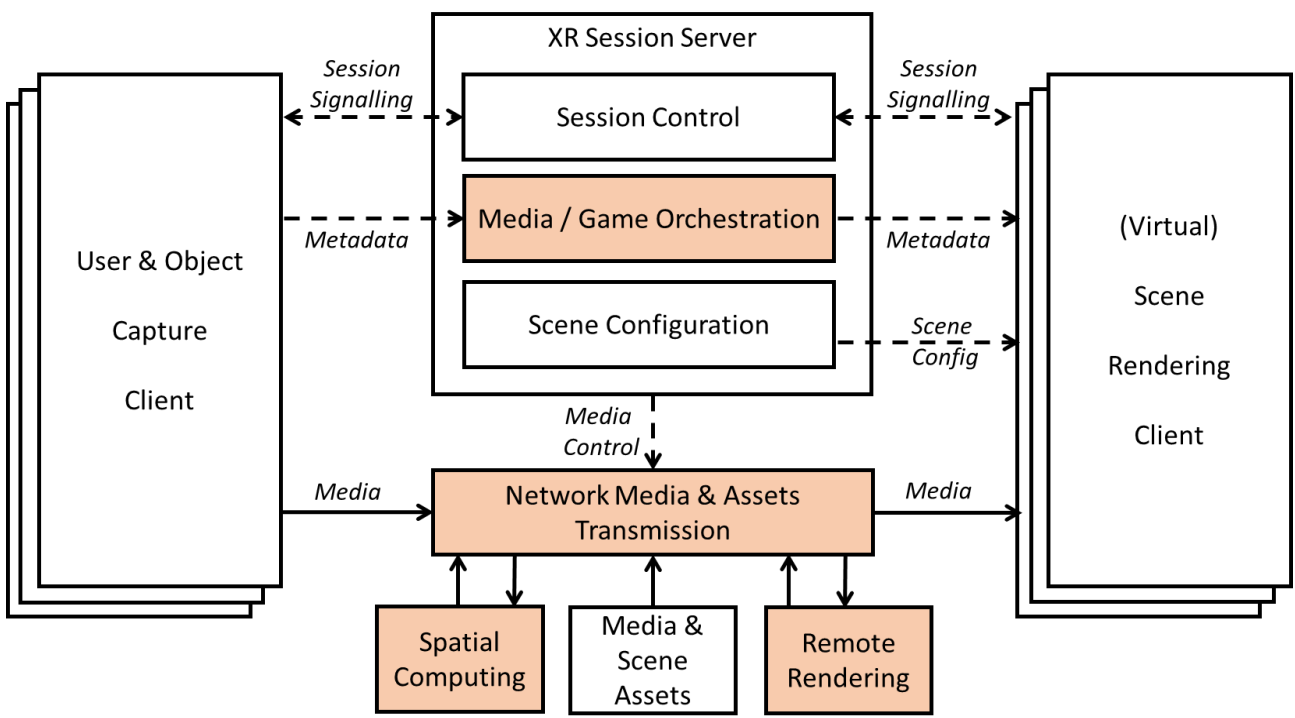

Figure 1 – Generic Multi-User XR Architecture

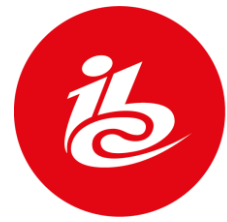

**ARCHITECTURE AND FUNCTIONAL BLOCKS**

With the strong demands and rapid technological developments in XR, systems and services are adapting accordingly. A comprehensive overview can be found in the 3GPP technical recommendation TR 26.928 (4). This recommendation includes multiple use cases and potential architecture solutions. However, in this work, multimedia aspects and game-centred use cases and architectures are still mostly studied separately. Figure 1 presents a more generic perspective of an XR architecture that supports both multi-user game and collaborative multimedia use cases.

Figure 1 is an extension of the "General architecture for XR conversational and conference services" (TR 26.928 (4), Figure 6.2.8-1) to include a more complete overview of different core XR functional blocks (including game engine flavours). Within Figure 1, we can identify 4 distinct new functional blocks that will be explained in the remainder of the paper:

I. Spatial Computing
II. XR interaction modalities
III. XR Immersive media formats & transmission
IV. XR Remote rendering

From these 4 blocks, the main difference towards the existing game architecture is to allow real-time multimedia and dynamic assets streaming. Integrating real-time video streams is still a challenge in most game engines. Furthermore, even though cloud gaming is already working well, many new requirements need to be addressed for complex spatial XR rendering (i.e., on low powered AR devices). Regarding common multimedia architectures, the biggest change is in the spatial aspects, being in a 3D environment with geometry is significantly different from the traditional 2D approach of media applications. Simply, any media needs the awareness of the real-world environment and manipulation of real-time media recordings of users or objects to reflect virtual rendering.

**SPATIAL COMPUTING**

One of the main differences of XR towards traditional media applications is the aspect of space (or in other words, 3D geometry). Computer games rapidly evolved from a 2D perspective towards 3D rendering. Thus, there are a lot of solutions existing to map towards XR. However, the focus of game engines is not much on real-time acquisition of multimedia assets rather than prerendered content and 3D assets. Sensing the user and his/her environment is exactly where Spatial Computing is closing the gap. Spatial computing as a term, however, is very generic and incorporates many aspects about sensing anything related to the locations and geometry of users and objects (5). Relating to XR, this means widely autonomous operations to understand the environment and activities of the users and thus building the main enabler for any AR technology. This could be for simply controller input interaction modelling or complex visual-based analysis like 3D mapping of the real-world space to blend objects into the environment (6). For this paper, we do not directly focus on the functional blocks a Spatial Computing component would offer or be composed of, but on the architectural implications and new metadata it required to enable Spatial Computing functionality.

For photo-realistic XR experiences, this also includes the Spatial Computing functionality to create 3D representations of users and objects. Either via special (e.g., RGB+Depth) video cameras that allow the creation of holographic media content (e.g., point clouds or meshes) or via prerendered 3D assets that are animated with the movements of the entities they represent (e.g., objects, humans, or animals). This mapping of real and virtual entities into the rendered or hybrid (blended with the real) environment is the key for any immersive media (or XR) application (7). Both cases create the need for new metadata, in particular how and where to place any entity and how to orchestrate and synchronize these entities across multiple end devices.

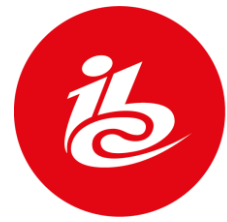

**Entity Placement & 3D Formats**

In multimedia, there is a long interest on how to compose, orchestrate, and render different types of media. One of the first approaches for a unified synchronized multimedia integration language (SMIL) (8) was standardised in 1998 by W3C[1] (now existing in version 3.0[2]). SMIL is a fully XML-based descriptive language to handle different media modalities like video, audio, text, etc. Mainly, to define media layout, timing, and synchronisation. SMIL is still supported in most browsers but current adaptation is limited to 2D content. A more recent approach to the positioning and allocation of media in XR, and more specifically spherical (e.g., 360-degree) video content, were specified by the MPEG Omnidirectional MediA Format (OMAF) (9). OMAF allows for a wide number of XR use cases, but it is not clear if it can cover all aspects of spatial and geometrical 3D contents. OMAF follows a multimedia centric approach and in this way can be considered separated from most game engine approaches. Each of the various game engines (the most popular for XR development being Unity and Unreal) follow their own respective proprietary mechanisms for object-to-scene composition while usually allowing export to different 3D file formats for interoperability. Many of such proprietary 3D formats exist, all with different feature sets to define 3D content, its position, and rendering properties. One particular format to mention is the Graphics Language Transmission Format (glTF[3]). GLTF is a royalty-free format that can both describe the composition of 2D and 3D media data, as well as directly compress assets in its binary format. Within the current version 2.0, glTF has already found a widespread adaptation in many 3D tools, game engines, and web browsers. It is however yet to be seen if it can be established as a main entity placement descriptive format for XR.

**Synchronisation and Media orchestration**

Regarding multi-user and distributed experiences, synchronisation and media orchestration need to be at the heart of the system. Thus, it comes as no surprise that this is a very active field of research with many books and research articles addressing different multimedia synchronisation and orchestration strategies, e.g., a recent overview of different optimisation strategies can be found in (10). One example of a synchronisation architecture is DVB-CSS (11), which compiles an extensive set of specifications to synchronize multiple media sources and many open-source implementation examples.

Not only from a multimedia perspective, but also from a game engine perspective, synchronisation is important. Every modern computer game engine has many mechanisms to optimise delays and reduce lags with various prediction mechanisms to allow object state synchronisation on a large scale (e.g., 100+ players in BattleRoyal games like PUBG and Fortnite). However, game engines usually only deal with very restrictive user interaction, predefined content and prerendered assets and might fall short on real-time captured and transmitted data.

---

[1] https://www.w3.org/TR/1998/PR-smil-19980409/

[2] https://www.w3.org/TR/SMIL/

[3] https://www.khronos.org/gltf/

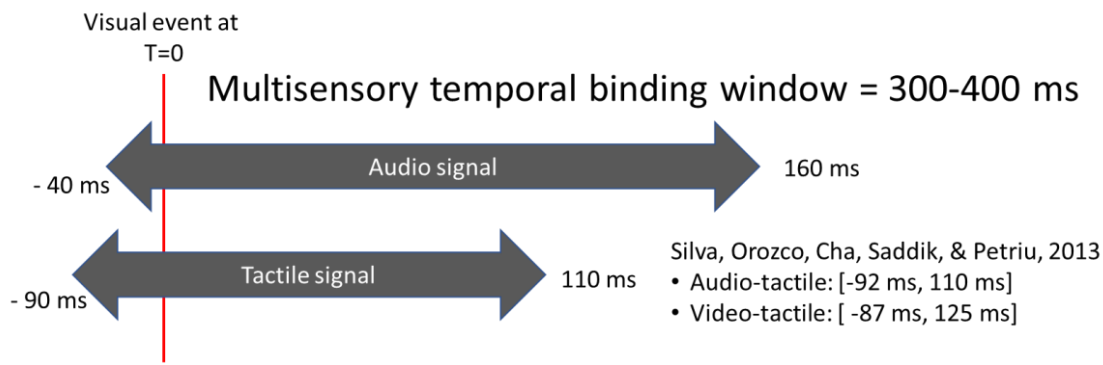

Figure 2 – Multimodal media synchronisation

Figure 2, shows the synchronisation requirements of the different modalities discussed in this paper, based on (12). Taking a visual event as a centre of the timeline, we expect that other interaction signals can appear earlier or later in time, without significant impact on the user experience. Therefore, allowing different boundaries of operation for an optimised transmission. Overall, XR has many different assets to build upon to orchestrate XR media assets, but with many new emerging interaction modalities, metadata, and transmission formats, existing solutions might need further adaptation.

**XR INTERACTION MODALITIES**

XR communication platforms allow for a wide range of interactivity, involving multiple modalities such as eye gaze, body/hand gestures, and speech. In recent years, many explorative studies (13) and experimentation (14) led to a better understanding of the various interaction methods enabling a more immersive environment. A multimodal platform takes advantage of the multiple human sensory channels in a virtual domain, allowing for diverse types of interactions, and therefore creating a better user experience.

**Eye tracking and eye gaze: Foveated Rendering**

Eye tracking, or gaze tracking, measures the gaze direction of the user. Almost all eye trackers work according to the same principles, near-IR light sources are used to illuminate the eye (illuminators), and high-resolution cameras identify the reflection of this light source on the cornea and in the pupil. A 3D model of the eye is then used to compute the reflection vector, which will be the gaze direction.

Eye gaze tracking is already widely utilized in many fields of research including VR/AR. When combining eye gaze direction with positional head tracking, one can determine the exact point the user is looking at in the real or virtual world. This feature has become quite relevant for the adoption of XR technology, given its contribution to system integration and user experience[4]. Concerning system integration, eye gaze tracking unlocks critical capabilities in the VR device such as foveated rendering, which can be used for better graphics performance, reducing the GPU load, and guaranteeing smoother frame rates, or increasing battery life (15). Regarding the user experience, eye-gaze can enhance the VR experience by constantly tracking the user's visual attention, which visual elements trigger certain responses and behaviours, and opens the possibility of social eye contact in virtual reality. For this reason, eye gaze tracking is rapidly becoming a standard feature in high-end VR headsets (16).

---

[4] https://blog.tobii.com/partnerships-in-xr-fundamental-to-mass-market-uptake-c30b0f655538

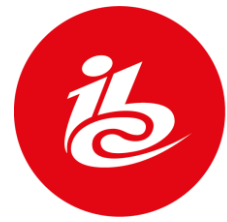

**HMD Removal for more immersive visual**

Social eye contact can dramatically improve the feeling of immersion in virtual meetings, both in professional and consumer applications, as the capture of nonverbal behaviour is considered essential to support collaborative interactions that are more like those in the real world (17). It allows users in VR to share their focus of attention and establish joint attention, which contributes to the feeling of immersion, and ultimately to social presence (18). Furthermore, by leveraging social eye contact in VR, avatar-based applications, or volumetric user capture applications with the HMD removed through AR, can become realistic with natural eye movements. Together with head-and-hand gestures, eye-gaze enhances the VR experience, applications become more intuitive, learning is simple, and the user experience is memorable.

Performing HMD removal comes at a cost, it often requires preprocessing and capture of the user prior to the experience (often introducing or prolonging an initialization step), such process can be rather computationally heavy leading to delays. To prepare the inpainting elements for the HMD removal, the user is first captured without the HMD by capturing various facial expressions. During the immersive experience, the location of the HMD in the image is then determined, and the points or pixels of the HMD location are replaced either by the precapture (19) (20) or by reconstructing the facial expression using a model-based approach (21).

**Haptic feedback for touch modality**

Haptics is a technology that stimulates the sense of touch and motion, simulating sensations to a user by reproducing the effects felt on a physical object. From the perspective of a user, "haptics" can be divided into haptic input and haptic feedback. Haptic input can be simply regarded as the captured data on the position of body parts, i.e., relying on motion sensors.

Haptic feedback can be divided into kinaesthetic and tactile feedback. Kinaesthetic feedback provides detailed proprioceptive "feeling" information, for example, on held objects, by multiple sensors in the body, like muscles, tendons, and joint angles. Information such as force, torque, position, and velocity are received by the skeleton, muscles, and tendons in the body (22). Tactile feedback is a subset of haptic feedback and is the feeling one gets from the mechanoreceptors in the surface tissue of the skin. These can sense things like vibration, pressure, temperature, texture, and touch[5]. Besides electric vibratory actuators, less common tactile actuators are piezoelectric benders, air puffs, and electrodes (23). Vibro-tactile information can be based on amplitude, frequency, timing, and location. Detection thresholds differ between individuals, suggesting the need for users to be able to adjust intensity (23).

When haptic technology is applied to elicit, enhance, or influence the emotional state, it is regarded as "affective haptics". Focusing on tactile feedback will provide an increase in the sense of immersion without having to invest in complex devices that do not necessarily improve immersion. However, while not focusing on kinaesthetic feedback, a kinaesthetic illusion can be created by stimulating specific body parts, e.g., vibrotactile actuation on tendons. The user then has the illusion that a limb (or other body part) is in a certain position. Users can have various levels of experience with XR and tactile displays, which influences the complexity of data that can be presented and perceived by the user. Users can be trained to interpret tactile data, as it is the case with modern mobile phones; different distinguishable vibrotactile patterns can represent different actions.

---

[5] https://manjubhat.wordpress.com/2014/04/06/tactile-vs-haptic-feedback/

Not every form of tactile feedback needs to be "social touch" (24) to increase immersion in VR. The illusion of interacting with objects without any social aspect can increase the immersion of the experience. Non-social (or regular) touch in any XR system will be a prerequisite for the successive development of "social touch".

Because the tactile displays need to convey their data to the mechanoreceptors of the body, it is necessary to note that the human body has a nonconstant spatial resolution. The skin is more sensitive to different types of data with varying spatial resolution in different places; for example, the back is less sensitive than the fingertips. This might mean that it is possible to display more complex information patterns to, e.g., fingertips than other body parts (24). Therefore, it is important to choose the most suitable display (e.g., a glove instead of a vest), depending on the use case.

**XR IMMERSIVE MEDIA FORMATS & TRANSMISSION**

Traditionally, multimedia transmission was centred around the video, audio, and text modalities. However, recent trends in immersive media led to evolutions both in the types of modalities supported and in the form of existing modalities. In this section, we focus on the evolution of the video modality to accommodate immersive XR applications, as well as the introduction of new modalities, using as an example tactile data. Video research has already significantly improved the quality of 2D video delivery (supporting HDR, 8K+ etc.). On the audio side, many improvements have been done both for cinematic home experiences and immersive media, e.g., via MPEG-H Object-based 3D audio, Dolby Atmos, Ambisonics. In this paper, we focus on the video aspects and the next step on moving from 2D to 3D immersive media formats, also referred to as holographic video. Additionally, we present the example of haptics as a new modality which is one of the emerging non-AV data formats that is expected to significantly affect the level of immersion.

**Holographic video**

Regarding holographic video, there are several approaches in facilitating spatial expressiveness, ranging from 3D model-based approaches that borrow elements from the gaming domain, in that it expresses 3D models and their spatiotemporal relationships, to video-first approaches, where video concepts are extended with functionality to provide a sense of volume on the rendered media.

With the 3D-based approach, traditionally static content (e.g., meshes) are given a time dimension to be suitable for consumption as holographic video content. Currently, Point Clouds are preferred over meshes for such purposes since it can directly express the content as captured (e.g., via LiDAR cameras). Point Cloud codecs for volumetric video are already being adopted and standardized by the industry, for example, the Video-based Point Cloud Codec (V-PCC a.k.a. V3C), currently developed by MPEG, that uses techniques known from the video encoding domain to compress the point clouds and their associated colour information (25). For mesh-based approaches, there is activity, but still mostly within academia, like, for example, using mesh compression for spatial information in conjunction with video compression for matching textures (26).

Video-based approaches differ from the 3D-based in that the holography is expressed with consideration to the differences of the frames between viewing angles, instead of attempting to represent the whole scene in 3D. Early examples of such attempts include multiview video, that packages frames from different viewing angles within the same container and achieves compression efficiency via prediction, not only in the time domain (as with traditional video), but also between different views (e.g., MPEG MVC (27)). Depending on the rendering device, this can result in a holographic experience, or in a more traditional approach of multiview video to recreate or enhance a 3D capture representation (28).

Finally, "immersive video" compression and transmission is the most recent development in video holography. "Immersive Video" is an umbrella term for holographic video that does not fall under the aforementioned categories. This can be interpreted as light field video that tries to recreate a scene not only based on the light colour, but also on the light direction, giving flexibility to the reconstruction, like the layered scene capture, by Google (29). MPEG is also working on an activity entitled Immersive Video (MPEG-MIV) that uses some preexisting elements (e.g., the video-encoded atlases of V-PCC) combined with new features (e.g., applying pruning prior to generating patches) to achieve a similar result (30).

**Tactile data: Patterns**

Several levels of information complexity can be achieved on tactile displays by creating complex patterns. When these patterns are not intuitive (i.e., already learned by nonvirtual interaction with the world) or *self-explaining*, the user needs to map the combination of stimuli to the intended meaning (i.e., training is required).

Given enough training, a user will be able to distinguish large amounts of data through its haptic feedback devices. However, this *does* take training, which humans are very well adapted to this (for example, any video game controller can be regarded as nonintuitive and with minor training most users will interact without problems). Complex patterns thus represent a *shared symbolic meaning.*

Patterns can be created by variations in multiple factors, depending on the modality that is used. In brief, the capacity to create patterns depends on the method and resolution in which the changes in stimuli are perceived. By varying the several factors of stimuli, complex patterns can be perceived and interpreted.

**Location:** probably the most trivial factor. The human body has a very high resolution for detecting *where* a stimulus occurs. This varies on the body, e.g., the hand has a higher spatial resolution compared to the back. This might also vary between people.

**Duration:** the time a tactile stimulus is active can vary. The human body can detect very short stimuli. Duration can be used to encode various meanings.

**Frequency:** Vibrotactile actuators can vibrate with different frequencies, although actuators might be optimized for a small frequency band.

**Intensity:** Intensity capabilities may vary. For vibrotactile actuators this normally is related and proportional to frequency, however other modalities like electrical stimuli can vary independently.

**Intermittence:** Multiple stimuli within a certain timespan can encode patterns, thus this factor includes a clear temporal dimension. Short bursts of stimuli or repeating simple patterns can represent a specific meaning.

**Complex combination:** Combining multiple factors of the above can create very complex tactile patterns. This may increase the amount of symbolic data but will presumably require more *training time* and *attention* from the user to interpret.

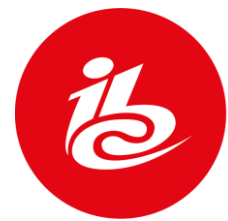

| Modality | Usage | Information |
|---|---|---|
| Vibration | Mobile phones, smart watches | Patterns of information might be used for *kinaesthetic illusion.* |
| Temperature (heat/cooling elements) | Material simulation | Heat flux makes one material feel colder than another while both have the same absolute temperature. |
| Pressure | Experimental | May even be established with focused ultrasound without physical contact. |
| Electro-stimulus | Surface properties of virtual objects | Can convey information from a subtle tinkling sensation to pinch-like stimuli. |

Table 1 – haptic / tactile interaction modalities

**XR REMOTE RENDERING**

XR developments often require repurposing of existing technology such as user interfaces, interaction methods, and 3D rendering; technologies which are commonly used by the gaming industry. Since the release of the computer game DOOM in 1993, which started the transition to 3D graphics, games have become increasingly larger and more advanced. The virtual worlds created are usually complex, with an increasing amount of geometry, number of textures, complexity of lighting, animation, and physics. Additionally, these worlds are dynamic and change according to the user's progression and interaction. To arrange the assets and models in a virtual world and improve the efficiency of rendering, game engines have integrated spatial data structures. The most common are scene graphs, spatial partitioning, and bounding volume hierarchies (BVHs). However, each game engine or 3D program has their own implementation of these spatial data structures. Further to the rendering engine, different end device hardware also impacts the rendering chain and results in multiple techniques and solutions (31). Thus, when sharing virtual worlds across users and platforms, it is beneficial that a standard file format can be used. For the description of individual 3D objects, there are already several formats (e.g., 3DS, OBJ, and FBX). For full scene description and animation, Kronos Group has developed the glTF format, which gained adoption and is currently predominantly used in web-based engines. With many different XR devices available, a standardized format of scene description will ease deployment and sharing of content across engines and architectures.

**Game engines in media production**

With their increasing functionalities, game engines are expanding to other media platforms like television. Unreal Engine is used by TV shows like The Mandalorian and West World to generate in-camera VFX, which is a new method for creating real-time visual effects during live action film recording (3). The technique requires low latency camera tracking through optical -, feature -, and inertial tracking methods, and realistic rendering of the VFX according to the camera's position. These tracking and rendering techniques can also be applied to XR applications, for this method of creating a real-time virtual environment is already by itself a form of extended reality. Creating such a real-time graphics system demands substantial processing power, especially when trying to achieve accurate lighting simulations with ray-tracing algorithms. A new generation of graphics cards with specialized ray-tracing cores by Nvidia and AMD makes this possible, and games like Cyberpunk 2077 and Metro Exodus show that real-time raytracing is now widely accessible. However, these games are based on prerendered 3D models and virtual worlds that can be fully controlled. The challenge in XR applications lies in many unknowns, e.g., location, lighting, occlusion, orientation, and interaction.

**Remote Rendering / Media streaming**

Even though powerful graphics systems are becoming more common, they are still rather sizable and unsuitable for handheld devices or headmounted systems. Cloud - and gaming companies have been looking at methods to step into this void and create remote gaming and rendering solutions. Nvidia's CloudXR[6] provides a platform for streaming XR content to untethered devices and Microsoft aims to offer similar solutions on their Azure[7] cloud products. CloudXR allows the use of advanced graphics techniques on remote servers to be rendered on mobile XR systems. A virtual world in a game engine is running on a server, with virtual cameras in the scene providing the video stream for the mobile devices. A separate connection to the server contains the XR devices' position and orientation. Compression algorithms and low network latency are crucial to the effective implementation of split rendering, but when done correctly can provide new opportunities for (photo)realistic XR experiences. One example of such an architecture can be found in (32), which allows device-independent remote rendering of complex photo-ralisitc 3D content. There is a clear benefit and necessity for remote (split) rendering to support light weight AR devices with limited battery power. However, one main challenge is in balancing the shift of rendering (and other processing) resources with the energy needed to transmit real-time 3D assets between the end device and the cloud (or edge).

**CONCLUSIONS AND OUTLOOK**

The main building blocks relevant to a XR system were outlined in this paper, both from a game engineering and multimedia streaming system perspective. Our overview provides the foundation for a complete multi-user and multisensory XR architecture, where new metadata and media orchestration are key for achieving complex interactions between users. We have shown throughout each fundamental block that gaming engines and multimedia orchestration require a closer synergy, optimizing processes at various levels.

Such synergy should also be reached in standards to create more coherent formats and interoperable systems. With the emergence of new media formats (i.e., holographic content and haptic data) standards should provide a global approach that integrates both worlds.

Functional components as well as the XR architecture are detailed in this paper, where modular aspects are considered in both game engines and multimedia streaming. XR is becoming mainstream for remote communication and is increasingly used in various use cases. It is crucial to consider its full capability in both game engines as well as multimedia domain to fulfil the high demands and ever stringent requirements on social perception and multi-user aspects.

---

[6] https://www.nvidia.com/en-us/design-visualization/solutions/cloud-xr/

[7] https://augmentit.ch/what-is-azure-remote-rendering-and-why-is-it-needed/

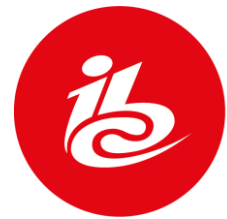

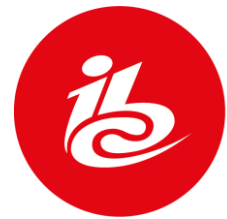